\documentstyle[prb,aps]{revtex}
\begin{document}
\draft
\title{Quasi-Hamiltonian Equations of Motion for Internal Coordinate Molecular
Dynamics of Polymers}
\author{Alexey K. Mazur}
\address{Laboratoire de Biochimie Th\'eorique, CNRS UPR9080 Institue de
Biologie Physico-Chimique 13, rue Pierre et Marie Curie, Paris,75005,
France$^1$}
\address{Pacific Institute of Bioorganic Chemistry, Russian Academy of
Sciences Vladivostok, 690022, Russia}
\date{\today}
\maketitle

$^1$Present address. Email: alexey@ibpc.fr

\begin{abstract}
Conventional molecular dynamics simulations macromolecules require long
computational times because the most interesting motions are very slow
compared with the fast oscillations of bond lengths and bond angles that
limit the integration time step. Simulation of dynamics in the space of
internal coordinates, that is with bond lengths, bond angles and torsions as
independent variables, gives a theoretical possibility to eliminate all
uninteresting fast degrees of freedom from the system. This paper presents a
new method for internal coordinate molecular dynamics simulations of
macromolecules. Equations of motion are derived which are applicable to
branched chain molecules with any number of internal degrees of freedom.
Equations use the canonical variables and they are much simpler than
existing analogs. In the numerical tests the internal coordinate dynamics
are compared with the traditional Cartesian coordinate molecular dynamics in
simulations a 56 residue globular protein. For the first time it has been
possible to compare the two alternative methods on identical molecular
models in conventional quality tests. It is shown that the traditional and
internal coordinate dynamics require the same time step size for the same
accuracy and that in the standard geometry approximation of aminoacids, that
is with fixed bond lengths, bond angles and rigid aromatic groups, the
characteristic step size is 4 fsec, that is two times higher than with fixed
bond lengths only. The step size can be increased up to 11 fsec when
rotation of hydrogen atoms is suppressed.
\end{abstract}

\section{Introduction}

Computer simulations of molecular dynamics (MD) have become an invaluable
tool applied to numerous problems in chemical physics \cite{Allen:87} and
biophysics \cite{Brooks:88}. Normally molecular movements are simulated by
the method of point particles \cite{Hockney:81}, i.e. motion of each atom is
computed by Newton's equations while molecular structures are maintained by
harmonic potentials that keep bond lengths and bond angles close to their
standard values. The maximum time step with which Newton's equations can be
numerically integrated is limited to rather small values (1 fsec for systems
containing explicit hydrogen atoms), in order to reproduce bond-length
oscillations accurately. This time step is annoyingly small for simulations
of macromolecules, notably biopolymers. The unique molecular properties of
biopolymers result from specific concerted atom movements, conformational
transitions, which can occur over a very broad time range: from picoseconds
to seconds. Because of the necessarily small time step, interesting motions,
for the most part, remain beyond the reach of even the fastest computers.
This presents the central problem in methodology in the field.

Twenty years ago Ryckaert et al \cite{SHAKE:} proposed an approach to this
problem which is now called constraint dynamics. They found a simple and
efficient way to impose holonomic constraints upon atom-atom distances in a
system of point particles governed by Newton's equations. Their well-known
algorithm, SHAKE, and a few other techniques based upon the same idea
\cite{Rattle:,EEM:} are now often used to fix bond lengths,
increasing the time step limit to 2 fsec for systems with explicit hydrogens.
Constraint dynamics has been extensively reviewed
\cite{Allen:87,Ciccotti:86,Tildesley:93}. These simple algorithms can
be applied to any molecules regardless of their size, chemical structure etc.
However, although it seemed initially that the method could fix not only bond
lengths but also, by triangulation, could constrain bond angles, dihedral
angles, rigid planar groups etc. \cite{SHAKE:}, this was found to be true
only for very small molecules. In large complex polymers like proteins even
bond angles cannot be fixed in this way \cite{vanGunsteren:82}.

The intrinsic limitation of constraint dynamics can be qualitatively
understood from the underlying physical model. Imposing a distance
constraint implies that a reaction force is introduced which is applied
along a line joining two atoms. Reactions must be calculated at each time
step to balance all other forces in the system, which for large molecules
presents a difficult problem because they are all coupled and form a system
of algebraic equations solved by iterations. With only bond lengths to
hydrogens fixed reactions are coupled in small groups and iterations
converge rapidly. With all bond lengths fixed reactions are coupled globally
and looping becomes possible. In general, in this case, the convergence is
not guaranteed \cite{Barth:95}, but, in practice, it remains acceptable. When
bond angles are triangulated coupling is much stronger and convergence
becomes too slow. Efforts to overcome these limitations continue
\cite{Durup:91}; in particular, it was shown that an exact non-iterative
calculation of reactions is possible, in principle \cite{Turner:95}, but,
so far, this is practical only for water molecules \cite{Miyamoto:92} and
for linear unbranched chains in which case the problem can be reduced to
inversion of a banded matrix \cite{EEM:,Gronbech:94}.

Despite these difficulties it remains desirable to model chain molecules
within the so-called standard geometry approximation \cite{Go:69}, that is
with standard fixed conformations of individual chemical groups and with
rotations around single bonds as the only motion considered. Rotatable
torsion angles completely define molecular conformations and present a
convenient minimum set of generalized coordinates which are directly used in
static conformational analysis and Monte-Carlo simulations of polymers. In
principle, MD simulations in the space of generalized, rather than
Cartesian, coordinates are also possible by using one of the appropriate
general formalisms of the classical mechanics. In this way all limitations
imposed by the fast vibrations of bond lengths, bond angles etc. would be
eliminated automatically. This approach was first applied to liquid butane
with one internal degree of freedom \cite{Ryckaert:75} and soon after Pear
and Weiner pioneered simulations of linear chains of up to 15 bonds based
upon the Newton-Euler equations of rigid body dynamics \cite{Pear:79}.
Interest in this line of research next decreased because of its complexity
and because it initially appeared that constraint dynamics could solve all
problems, but the difficulties encountered have led to a revival of work.

To date several research groups have reported attempts to find a suitable
technique for MD simulations in generalized coordinates \cite{Pear:79,%
Perram:88,BKS0:,BKS1:,BKS2:,Dor-int:,Gibson:90,Jain:93,Mathiowetz:94,%
Rice:94}. Both methods of rigid body dynamics \cite{Pear:79,Perram:88,%
Jain:93,Mathiowetz:94,Rice:94} and the Lagrange-Hamilton formalism
\cite{BKS0:,BKS1:,BKS2:,Dor-int:,Gibson:90} were
employed. These methods often have little in common in their analytical
formulations, but they all may be reasonably referred to as {\em internal
coordinate molecular dynamics} (ICMD). The full set of internal coordinates
includes not only torsions, but also bond lengths and bond angles, and in
some approaches these variables may also be free, if desired. The term {\it
internal coordinate molecular dynamics} underlines the main distinction of
these approaches from conventional MD, i.e. they all consider molecular
motion in the space of generalized internal coordinates rather than in the
usual Cartesian coordinate space.

None of the ICMD techniques has yet been sufficiently well developed to
reach its main goal, that is, to compute long-duration macromolecular
trajectories with acceptable accuracy but at a lower cost than the classical
MD with bond length constraints. Compared with the pure Newtonian dynamics
there are two obvious problems with ICMD. The first is the derivation of
equations of motion for large molecules. This task seemed too complicated at
the beginning \cite{SHAKE:}, but several acceptable solutions have been proposed, which
represent the main achievement of the early reports on the
subject \cite{Pear:79,Perram:88,BKS0:,BKS2:,Rice:94}. The
second problem is the cost of additional calculations to be performed in
order to obtain generalized accelerations. This must be low enough to be
compensated by an increase in the step size. Among these calculations the
necessity to invert the mass matrix of the system is an evident obstacle.
This matrix is diagonal for the Newton's equations but for internal
coordinates it is a full positive definite matrix, therefore, a direct
inversion scales as $O(n^{3})$ with the number of degrees of freedom and this
quickly becomes impractical when n exceeds ~100. Fortunately, a solution of
this problem was found some time ago in robot mechanics
\cite{Vereshchagin:74,Featherstone:87,Rodriguez:87,Rodriguez:92},
namely, there are several recursive techniques that allow one to compute
exact generalized accelerations for $O(n)$ calculations if the system can be
treated as a tree of articulated rigid bodies. Two such algorithms have been
employed recently \cite{Mathiowetz:94,Rice:94} for ICMD and it was shown
that, as in Newtonian dynamics, the cost of the time step can be made close
to that of the evaluation of forces \cite{Mathiowetz:94}.

In spite of this progress the capabilities of ICMD still are not obvious
because it is unclear how much the time step can be increased in practically
interesting cases. The uncertainty is caused by two different reasons.
First, biopolymers with only torsion degrees of freedom are very complex,
heterogeneous and essentially unharmonic systems in which it is usually
impossible to distinguish the fastest motions {\em a priori} by intuition.
For instance, in simulations of small peptides time steps can be increased up
to 15-20 fsec \cite{Dor-int:,Mathiowetz:94,Dor-enk:,Pet-La:} and
one might expect that a considerable increase of the step size should be
possible for larger molecules as well. Surprisingly, the first simulations
of torsion dynamics of proteins reported recently
\cite{Mathiowetz:94,Rice:94} encountered a step size barrier of only 2 fsec,
the same as in the constraint dynamics with fixed bond lengths. This unusual
effect of molecular size still have no clear explanation. Most of the
uncertainty, however, results from unresolved numerical problems involved in
integration of ICMD equations of motion. Their analytical form does not give
a possibility to employ common symplectic MD integrators and that is why
various general-purpose predictor-corrector schemes were usually employed.
It is well-known, however, that in MD applications such algorithms tend to
loose stability with lower step sizes. In addition, it was shown that they
poorly conserve momenta when applied to equations of motion in generalized
coodrinates \cite{Dor-int:}.

This study presents an attempt to overcome the numerical difficulties in
ICMD reported in ref.\ \onlinecite{Dor-int:}. These difficulties originate
from the general form of the equations of motion which depends only upon the
choice of coordinates, and so they could not be overcome by simple remedies.
That is why an essentially new analytical formulation of ICMD has been
developed with new equations which are essentially Hamiltonian, but have a
slightly different form. Surprisingly, they turned out to be much simpler
than in other analytical ICMD formulations. Numerical integration in the
space of conjugate Hamiltonian variables eliminates all earlier problems
associated with poor conservation of momenta \cite{Dor-int:} and results
in trajectories as stable as those obtained with Newtonian MD.

This paper principally presents an analytical background, but includes
representative tests in order to obtain reliable estimates of the step size
in a few important simulation modes and to make sure that there are no
hidden numerical problems that can undermine the method. It turns out that
for protein torsion dynamics with all explicit hydrogens the optimal time
step is 4 fsec, two times larger than with only bond lengths fixed and the
same as that for rigid water simulations \cite{Fincham:92}. This value is
probably determined by fast collisions and rotations of hydrogen atoms
because without explicit hydrogens or with weighted inertia of hydrogen-only
rigid bodies the possible time step is beyond 10 fsec. The computer cost of a
time step for the torsion dynamics is similar to that of usual constraint
dynamics with fixed bond lengths. It is, therefore, a method of choice for
long simulations in which the accuracy of the standard geometry
approximation is acceptable. It should be also stressed that ICMD
qualitatively differs from the constraint dynamics in its efficiency with
respect to the number of constraints: additional constraints in ICMD result
in an increased speed of calculations.

\section{Derivation of Equations}

Let ${\bf r}_\alpha $ denote the position vector of atom $\alpha $. The
conformation of the molecule is specified by the set of $N$ such vectors
\{${\bf r}_\alpha \}$. Kinetic energy
\begin{equation}
K=\sum_{\alpha =1}^N\frac{m_\alpha{\bf\dot r_\alpha}^2}2
\end{equation}
where overdot denotes time derivative; the potential energy
\begin{equation}
U=U(\{{\bf r}_\alpha \})
\end{equation}
and the Lagrangian
\begin{equation}
L=K-U
\end{equation}
Assuming that the set of generalized coordinates is defined
$\{\theta_i\}\ i=1,...n$, so that
\begin{equation}
{\bf r}_\alpha ={\bf r}_\alpha \left( \{\theta _i\}\right)
\end{equation}
the Lagrangian equations of motion are
\begin{equation}
\frac d{dt}\frac{\partial K}{\partial \dot\theta _i}-\frac{\partial K}{\partial
\theta _i}=-\frac{\partial U}{\partial \theta _i}
\end{equation}
Now consider how molecular conformation is specified by bond lengths, bond
angles and dihedrals. If in Fig.1 the Cartesian coordinates are given for
atoms $\alpha$, $\beta$ and $\gamma$, ${\bf r}_\delta$ is computed from
${\bf r}_\alpha$, ${\bf r}_\beta$ and ${\bf r}_\gamma$ given the values of
dihedral $\omega$, planar angle $\varphi$ and bond length $l$
\begin{equation}
{\bf r}_\delta ={\bf r}_\delta({\bf r}_\alpha,{\bf r}_\beta,{\bf r}
_\gamma,\omega,\varphi,l)
\end{equation}
Similarly, ${\bf r}_\varepsilon$ is computed from ${\bf r}_\beta$, ${\bf r}
_\gamma$ and ${\bf r}_\delta$, and so on. This procedure needs global
coordinates of the first three atoms for initialization, but for the moment
we skip this point to consider it later. The above process is not just an
algorithm for computing \{${\bf r}_\alpha\}$ from $\{\theta_i\}$, but also
a definition of a specific ordering of atoms and internal coordinates in the
molecule. This ordering is indispensable for transformation of $\{\theta
_i\}$ into \{${\bf r}_\alpha\}$; it is implicit in Eq.(4) but is revealed
when derivatives $\partial{\bf r}_\alpha/\partial\theta_i$ need
to be computed. For example, when dihedral $\omega$ in Fig.1 is varied then
by definition ${\bf r}_\alpha$, ${\bf r}_\beta$ and ${\bf r}_\gamma$ are
not affected while ${\bf r}_\delta$ moves, as well as the following atoms.
The infinitesimal displacements of atoms in response to variation of
internal coordinates correspond to the motion of a tree: the very first
three atoms form the base, internal coordinates close to the base move
almost the whole molecule while variables in a specific branch move only the
upper part of that branch. Of course, there is no distinction between the
base and the tip of the tree in the real motion: when the base gets its six
degrees of freedom it becomes equivalent to any of the branches.

Let $D_i$ denote the set of atoms in the molecule that move when internal
coordinate $\theta _i$ is varied. Than we have
\begin{equation}
\frac{\partial K}{\partial\theta_i}=\sum_{\alpha\in D_i}m_\alpha {\bf\dot r}
_\alpha\frac{\partial {\bf\dot r}_\alpha}{\partial\theta_i}
\end{equation}
\begin{equation}
\frac{\partial K}{\partial\dot\theta_i}=\sum_{\alpha\in D_i}m_\alpha
{\bf\dot r}_\alpha\frac{\partial{\bf\dot r}_\alpha}{\partial\dot\theta_i}=
\sum_{\alpha\in D_i}m_\alpha{\bf\dot r}_\alpha\frac{\partial{\bf r}
_\alpha}{\partial \theta_i}
\end{equation}
and Eq.(5) gives
\begin{equation}
\sum_{\alpha\in D_i}m_\alpha{\bf\ddot r}_\alpha\frac{\partial{\bf r}_\alpha}
{\partial \theta_i}=-\frac{\partial U}{\partial \theta_i}
\end{equation}

In order to make the following step it is necessary to assume that there are
no loops in the molecule. Internal coordinate description of flexible loops
is a separate problem which needs special treatment, but without loops the
whole set $D_i$ moves as a single rigid body if only $\theta_i$ is varied.
If $\theta_i$ is a bond lengths $D_i$ is translated along the bond so that
for any atom
\begin{equation}
\delta{\bf r}_\alpha={\bf e}_i\delta\theta_i
\end{equation}
where ${\bf e}_i$ is the unit vector of the bond. If on the other hand
$\theta_i$ is an angle or a dihedral, rotation of $D_i$ occurs:
\begin{equation}
\delta{\bf r}_\alpha={\bf e}_i\times\left({\bf r}_\alpha-{\bf r}
_i\right)\delta\theta_i
\end{equation}
where the cross denotes a vector product, while the unit vector ${\bf e}_i$
and the position vector ${\bf r}_i$ specify the axis of rotation.

For a translational variable Eq.(9) gives
\begin{equation}
\sum_{\alpha\in D_i}m_\alpha{\bf\ddot r}_\alpha\frac{\partial{\bf r}_\alpha}
{\partial\theta_i}={\bf e}_i\sum_{\alpha\in D_i}m_\alpha{\bf\ddot r}_\alpha=
{\bf e}_i{\bf\dot P}_i=-\frac{\partial U}{\partial\theta_i}
\end{equation}
where ${\bf P}_i$ is the total momentum of the set of atoms $D_i$. It is
straightforward to show that, in the case of a translational variable,
$\partial U/\partial\theta_i$ equals to the sum of the forces
applied to atoms of set $D_i$ projected upon vector ${\bf e}_i$, and so Eq.
(12) is nothing but the corresponding projection of the Newton equation.
For a rotational variable we have
\begin{equation}
\sum_{\alpha\in D_i}m_\alpha{\bf\ddot r}_\alpha\frac{\partial{\bf r}_\alpha }
{\partial\theta_i}={\bf e}_i\sum_{\alpha\in D_i}\left({\bf r}_\alpha-
{\bf r}_i\right)\times\left(m_\alpha{\bf\ddot r}_\alpha\right)={\bf e}_i
{\bf\dot M}_i=-\frac{\partial U}{\partial\theta_i}
\end{equation}
where ${\bf M}_i$ is the total angular momentum of atoms of set $D_i$ around
the fixed point in space given by the current value of vector ${\bf r}_i$.
Again it is easy to show that, for a rotational variable, $\partial U/
\partial\theta_i$ equals the projection upon ${\bf e}_i$ of the sum of
the torques around point ${\bf r}_i$ applied to atoms of set $D_i$, and
therefore Eq.(13) is a projection of the Newton equations for the torques.

In order to shorten the notation let us introduce indicator $s_i$ such that

\begin{equation}
s_i=\left\lbrace\begin{array}{l@{\ \ if\ \theta_i\ -\ }l}
        1 & rotational\\
        0 & translational\\
        \end{array}
  \right.
\end{equation}
and unite Eqs.(12) and (13)
\begin{equation}
s_i{\bf e}_i{\bf\dot M}_i+\left(1-s_i\right){\bf e}_i{\bf\dot P}_i=
-\frac{\partial U}{\partial\theta_i}
\end{equation}
Now we need to transform Eq.(15) so that a full time derivative appears on
the left. We have
\begin{equation}
{\bf e}_i{\bf\dot P}_i=\frac d{dt}\left({\bf e}_i{\bf P}_i\right)
-{\bf\dot e}_i{\bf P}_i
\end{equation}
and
\begin{equation}
{\bf e}_i{\bf\dot M}_i=\frac d{dt}\left({\bf e}_i{\bf Q}_i\right)
-{\bf\dot e}_i{\bf Q}_i+{\bf P}_i\left({\bf e}_i\times{\bf\dot r}_i\right)
\end{equation}
where ${\bf Q}_i$ is the total angular momentum of set $D_i$ around moving
point given by vector ${\bf r}_i$. Although ${\bf M}_i$ and ${\bf Q}_i$
refer to the same physical quantity distinction between them
must be made because these are two different time functions. ${\bf M}_i$
is angular momentum around a fixed node; it equals to ${\bf Q}_i$ at the
given moment of time but it has a different time derivative:
\begin{equation}
{\bf\dot M}_i=\frac\partial{\partial t}{\bf Q}_i
\begin{array}{|l}\\{\bf r}_i=const\\\end{array}
\end{equation}
and, accordingly, the two values diverge immediately afterwards. The
Newton equation (13) is valid for ${\bf M}_i$ but not for ${\bf Q}_i$.

Substitution of Eqs.(16) and (17) into Eq.(15) gives
\begin{equation}
\frac\partial{\partial t}\left[s_i{\bf e}_i{\bf Q}_i+\left(1-s_i\right)
{\bf e}_i{\bf P}_i\right]=-\frac{\partial U}{\partial\theta_i}+s_i{\bf\dot e}
_i{\bf Q}_i-s_i{\bf P}_i\left({\bf e}_i\times{\bf\dot r}_i\right)+\left(
1-s_i\right){\bf\dot e}_i{\bf P}_i
\end{equation}
Now consider more carefully the term in the brackets on the left. By a
simple transformation we get
\begin{equation}
s_i{\bf e}_i{\bf Q}_i+\left(1-s_i\right){\bf e}_i{\bf P}_i=\sum_{\alpha
\in D_i}m_\alpha{\bf\dot r}_\alpha\left[s_i{\bf e}_i\times\left({\bf r}
_\alpha-{\bf r}_i\right)+\left(1-s_i\right){\bf e}_i\right]=\frac
{\partial K}{\partial\dot\theta_i}=\frac{\partial L}{\partial\dot\theta _i}
\end{equation}
i.e. this is nothing but the conjugate momentum corresponding to variable
$\theta_i$ and Eq.(19) is just one of the Hamiltonian equations. We have,
therefore
\begin{equation}
\sum_{k=1}^na_{ik}\dot\theta_i=s_i{\bf e}_i{\bf Q}_i+\left(1-s_i\right){\bf e}
_i{\bf P}_i
\end{equation}
where $\left\{a_{ik}\right\}$ are the coefficients of the mass matrix of
the system. Eqs.(19) and (21) present the resultant equations of motion to
be integrated by a computer. This is done in two steps. At first the r.h.s.
of Eq.(19) is evaluated and the time step is made for momenta. At the same
time momenta propagated from the previous step are substituted into the
r.h.s. of Eq.(21) and the linear system is solved by inverting the mass
matrix, which is done by using the recursive algorithm by Rodriguez et al
\cite{Rodriguez:87,Rodriguez:92}. This algorithm has been developed within
the context of the Newton-Euler approach to rigid body dynamics, where each
its step has a physical interpretation and corresponds to a special
decomposition of reactions in hinges between rigid bodies. Here these
interpretations are lost, but due to the same structure of matrix $\left\{
a_{ik}\right\}$ it still can be applied just as a formal mathematical
procedure. The generalized velocities thus obtained are used to make the time
step in generalized coordinates. A specific example of an integrator for Eqs.
(19) and (20) employed in the numerical tests reported below is detailed in
the Appendix.

Eq.(19) is rather simple and presents no computational problems. All terms
on the right are familiar functions of atom coordinates and atom velocities.
It is clear that in an unbranched chain molecule sets $D_i$ can be ordered
so that
\begin{equation}
D_1\supset D_2\supset ...\supset D_{n-1}\supset D_n
\end{equation}
therefore all ${\bf P}_i$ can be computed starting from the tip of the
chain, moving to the base and at the ith variable adding only the
contribution from subset $D_i/D_{i+1}$. In order to compute ${\bf Q}_i$
angular momenta of sets $D_i$ are first computed with respect to the zero of
global coordinates in the same way as momenta. After that ${\bf Q}_i$ are
computed by moving the ith node of rotation to ${\bf r}_i$ The total force
applied to atoms of set $D_i$ is computed similarly to ${\bf P}_i$ and the
total torque - similarly to ${\bf Q}_i$. It can be noted that the recurrent
calculations of derivatives of the conformational energy used in this field
for a long time \cite{BKS0:,Noguti:83,Abe:84} are nothing but the
above summation of forces and torques. Due to these algorithms computations
of forces in ICMD is in fact similar to the classical MD, contrary to the
common oppinion \cite{Jain:93,Mathiowetz:94,Rice:94}. Generalization of
the above procedure for a branched chain is straightforward. These
computations present only technical difficulties and there is no need to
detail them here. The computer time necessary to evaluate the non-potential
terms in Eq.(19) is negligible.

It seems reasonable to call Eqs.(19) and (21) Quasi-Hamiltonian because of
the meaning of the parameters involved. One should note that there is
already some confusion in the classification of equations of motion in
internal coordinates. It certainly has little sense to base such
classification upon the method used for the derivation of equations because
their final form depends only upon the choice of coordinates. For example,
by substituting Eq. (21) into Eq. (19) we can get the earlier equations for
generalized accelerations \cite{BKS0:} which are equivalent to any other
equations for accelerations of internal coordinates \cite{Jain:93}.
Inversely, Eq. (19) may be obtained by separating a full time derivative in
the corresponding equations for generalized accelerations derived by any
appropriate technique.

\section{Conservation of Momenta}

Let us consider now the generalized coordinates that position in space the
first three atoms in the molecule. These coordinates may be called external
and they affect positions of all other atoms, therefore ${\bf P}_1$ and $
{\bf Q}_1$ are the total momentum and angular momentum around a node which
is yet to be specified. In the absence of external field $\partial U/
\partial\theta_i=0$ therefore for external variables the r.h.s. of
Eq.(19) includes only inertial terms. All these terms become zero, however,
if variable $\theta _i$ can be defined so that the corresponding ${\bf e}_i$
and ${\bf r}_i$ are fixed in space. For instance, if the Cartesian
coordinates of the first atom are used as the first three translational
coordinates Eq.(19) gives
\begin{equation}
\frac d{dt}{\bf P}_x=\frac d{dt}{\bf P}_y=\frac d{dt}{\bf P}_z=0
\end{equation}
where ${\bf P}$ is the total momentum of the molecule. Thus, the
conservation of the total momentum appears to be explicitly present in
equations of motion and, more importantly, in their finite-difference
approximations. The same property is implicit in the Newton equations. On
the face of it, this might seem rather minor, but it appears to be somehow
related with the numerical stability with large step sizes. In the Newton's
case, among the principal first integrals of mechanics only the total energy
is affected by the numerical approximation errors, while the conservation of
momenta commonly is ``ideal'', that is it depends only upon the length of
the floating point numbers in the computer. In contrast, in the numerical
integration of standard ICMD equations for generalized accelerations all
first integrals are similarly affected by approximation errors. The
importance of this qualitative difference has been recognized in our earlier
study \cite{Dor-int:} where it was possible to significantly improve
the numerical stability with large step sizes by enforcing conservation of
momenta. By using Eqs. (19) and (20) ideal numerical conservation can be
ensured with an appropriate choice of the external coordinates, and we will
see below that in this way one manages to obtain the quality of numerical
trajectories comparable to that of classical MD.

The above considerations dictate the following procedure for positioning the
first three atoms in a molecule. Fig.1 shows the global coordinate frame
$xyz$ and the local frame $x^{\prime}y^{\prime}z^{\prime}$ that is bound with
the molecule. The origin of the local frame coincides with the first atom
(atom $\alpha$ in Fig.1). The second atom ($\beta$) always rests on axis
$O^{\prime}x^{\prime}$ and the third atom ($\gamma$) always rests in plane
$x^{\prime}O^{\prime}y^{\prime}$. The second atom moves along $O^{\prime
}x^{\prime}$ axis if its bond lengths (bond $\alpha$-$\beta$ in Fig.1) is
not fixed. Similarly, the third atom moves in $x^{\prime}O^{\prime
}y^{\prime}$ plane if bond $\beta$-$\gamma$ and/or bond angle $\alpha$-$
\beta$-$\gamma$ are variable. With these conventions, the six rigid body
degrees of freedom of frame $x^{\prime}y^{\prime}z^{\prime}$ with respect
to the global coordinates $xyz$ complement the internal degrees of freedom
in the molecule to the full set required.

Frame $x^{\prime}y^{\prime}z^{\prime}$ is considered to rotate around the
zero of the global coordinates, with variations of the first three
generalized coordinates, $\delta\theta_1$, $\delta\theta_2$ and $\delta
\theta_3$, corresponding to rotations around axes $x$, $y$ and $z$,
respectively. In this case $\dot\theta_1$, $\dot\theta_2$ and $\dot\theta_3$
obtained from Eq.(21) are the components $\omega_x$ $\omega_y$ and $\omega_z$
of the angular velocity of frame $x^{\prime}y^{\prime}z^{\prime}$. They are
used to integrate the kinematic equations for the quaternion that controls
orientation of frame $x^{\prime}y^{\prime}z^{\prime}$
\cite{Branets:73,Evans:77}. Vector ${\bf Q}={\bf Q}_1={\bf Q}_2={\bf
Q}_3$ is the total angular momentum around the global coordinate origin
therefore Eq.(19) guarantees conservation of the angular momentum. One can
note, however, that since the first atom is moved by variations $\delta
\theta_1$, $\delta\theta_2$ and $\delta\theta_3$ the Cartesian
coordinates of the first atom can no longer be taken as independent. However,
we can still select the three global translational variables so that
variations $\delta\theta_4$, $\delta\theta_5$ and $\delta\theta_6$
correspond to translations of all atoms along global axes $x$, $y$ and $z$.
In this case $\dot\theta_4$, $\dot\theta_5$ and $\dot\theta_6$ obtained from
Eq.(21) refer to $\upsilon_x$, $\upsilon_y$ and $\upsilon_z$ components of
velocity of the point that is rigidly connected with $x^{\prime}y^{\prime
}z^{\prime}$ frame, but instantaneously coincides with the zero of
coordinates as the center of rotation. The velocity of the first atom is then
computed from its position vector ${\bf r}_1$ and vectors ${\bf\upsilon}$
and ${\bf\omega}$. It is clear that in this case vector ${\bf P}={\bf
P}_4={\bf P}_5={\bf P}_6$ still is the total momentum of the molecule.

\section{Numerical Tests}

The formal validity of the analytical formulation of ICMD method described
in the sections above has been checked in numerous tests with biopolymers of
different size and chemical structure. These calculations clearly confirmed
the expected superiority of this new method over our previous formulation
\cite{BKS0:,BKS1:,BKS2:,Dor-int:}. In particular, results of the
tests obtained with enforced conservation of momenta \cite{Dor-int:}
have been considerably improved, especially as regards the magnitude of the
total energy drift (data not shown). The results of a few representative
calculations are shown in Fig. 2. Following many previous methodological MD
studies a single protein molecule in vacuum was considered here, and the time
step comparisons are made by using the conservation of the total energy to
access the accuracy of the numerical trajectory. The protein molecule chosen
is the immunoglobulin binding domain of streptococcal protein G \cite{pgb:}
(PGB, file 1pgb in the Protein Database \cite{PDB:}) which is a 56 residue
$\alpha $/$\beta $ protein subunit. PGB is somewhat larger than the commonly
used test proteins, but it does not have neither S-S bridges nor proline
residues, so that no complications connected with the treatment of the
flexible rings are involved. Two indicators of the conservation of the total
energy are used: the module of the drift and the relative fluctuation which
is computed as
$$
\delta =\frac{\sqrt{\left\langle \left( E-\left\langle
E\right\rangle \right) ^2\right\rangle }}{\sqrt{\left\langle \left(
K-\left\langle K\right\rangle \right) ^2\right\rangle }}
$$
where $\left\langle E\right\rangle $ and $\left\langle K\right\rangle $ are the
average total and kinetic energies, respectively. The value of the drift is
calculated by a linear regression during 1 psec intervals along the
trajectory and averaged.

In Figs. 2(a-d) the ICMD simulations are compared with the classical MD. For
the usual MD the popular package AMBER \cite{AMBER:} with force field
AMBER94 \cite{AMBER94:} was used, with distance dependent dielectric
constant, $\varepsilon =r$, and no truncation of non-bonded interactions. In
calculations with fixed bond lengths the SHAKE algorithm \cite{SHAKE:} was
employed with a tolerance of 10$^{-6}$. ICMD trajectories were computed by a
new program (DY) with the same force filed and all internal coordinates as
variables. Special care has been taken to ensure identity of the force filed
implementations in the two programs and identity of the starting states. The
equilibration was more or less standard and included minimization of the
crystal structure followed by a 12.5 psec run starting from Maxwell
distribution at 300 K with periodic temperature control and a step size of
0.5 fsec. After that the step size was reduced to 0.25 fsec and a short
trajectory of 150 fsec was calculated, with the final part stored and used
for generating initial data. The final point was used as the starting state
for a test trajectory of 10 psec, with initial half-step velocities taken at
appropriate time intervals from coordinates. These last preparations were
necessary to provide smooth starts of leapfrog integrators used in AMBER and
DY with different time step sizes. The same starting states were imported
from AMBER to DY where internal coordinates and generalized velocities were
computed from atomic coordinates and velocities. The test trajectory was
repeatedly computed with gradually growing step size first in Cartesian and
next in internal coordinates.

Figures 2(a,b) present results obtained in calculations with completely free
molecular models, i.e. with no constraints. The agreement between the two
sets of simulations is close to ideal. Note that the commonly used value of
the time step (1 fsec) corresponds to a relative fluctuation of $\delta $
=0.1, which can be reasonably used as a reference level of acceptable
accuracy. Results of analogous test with bond lengths constraints applied to
hydrogen atoms are shown in Figs. 2(c,d). Again one can note a remarkable
agreement between the two different methods. The relative fluctuation of 0.1
corresponds to the well-known time step size level of 2 fsec. A similar
agreement was observed in simulations with all bond lengths fixed, and it
should be added that absolute values of all energy components were very
close in all comparative tests (results not shown). These results
demonstrate that ICMD simulations are as accurate as traditional MD and that
both techniques suffer from similar time step limitations. Figures 2(a- d),
therefore, give an appropriate reference to which standard geometry ICMD
simulations may be compared.

The results of ICMD tests performed with the set of variables corresponding
to the standard geometry approximation are shown in Figs. 2(e,f). Compared
to Figs. 2(c,d) both the relative fluctuation and the drift of the total
energy are considerably reduced. The time step size corresponding to the
relative fluctuation of 0.1 is 4 fsec, that is two times larger than for
all- atom models with fixed bond lengths. This value is close to the
well-known step size limit for dynamics of rigid water molecules
\cite{Fincham:92}, suggesting that the limitation is probably due to the
fast rotation of hydorgen atoms in hydrogen-only rigid bodies like hydroxyl
or methyl groups. One can note that the observed increase in the step size
is rather moderate compared to earlier estimates obtained in simulations with
smaller molecules \cite{Dor-int:}. The apparent reason is that these
fast rotations should mainly occur in protein cores, and so the test molecule
must be sufficiently large and must include a large number of hydrogen-only
rigid bodies for these limitations to be detected.

Figures 2 (g,h) present an additional proof of the fact that, in the above
ICMD tests, the time steps are limited by rotation of hydrogens and
illustrates one simple possibility to get rid of this limitation. It is
clear that rigid body rotation can be slowed down by artificially increasing
its inertia. Imagine, for instance, that in ethane (CH$_3$-CH$_3$) both
carbon atoms are no longer point masses, but rather spheres that rotate
together with the neighboring hydrogens. Rotation of hydrogens will be
slowed down but the overall motion of the molecule will be perturbed only
slightly because atom masses do not need to be changed. This rather general
trick can be employed at all levels for balancing the time scales of
different motions in any hamiltonian system. Note, for instance, that a
similar approach is applied to electron degrees of freedom in the
Car-Parrinello method \cite{Car:85} and that the so called weighted
mass MD \cite{Pomes:90,Mao:91} is also based upon this idea. Results
presented in Fig. 5 (g,h) were obtained with point masses of atoms in the
rotation nodes of hydrogen-only rigid bodies replaced by spheres with the
same mass and fixed moments of inertia equal to 15 (atom mass unit)$\cdot$
\AA$^2$.
As a result, both the drift and the relative fluctuation are significantly
reduced compared to Fig. 2(e,f). The time step size that corresponds to the
relative fluctuation of 0.1 is about 11-12 fsec.

Some comparative data on the computer cost of simulations in AMBER and in DY
in different modes are given in the Appendix. Overall Table 1A shows that
ICMD is complementary to the traditional MD. As expected, for unconstrained
simulations and for the two usual modes of bond length constraints AMBER is
somewhat faster than DY. On the other hand, in the standard geometry
simulations the cost of a time step in ICMD is roughly the same as in
constraint dynamics with fixed bond length to hydrogens. Thus, the observed
increase in the step size is obtained essentially for no cost, and in terms
of computer time per picosecond ICMD gives the best score.

\section{Conclusions}

In recent years there has been a slow but steady progress in the development
of techniques for ICMD. As shown here, improved ICMD methodology now makes
possible simulations of torsion dynamics of biopolymers that fulfill the
same strict criteria upon accuracy and stability as classical MD. Overall
ICMD simulations within the standard geometry approximation are at least two
times faster than constraint dynamics.

It should be kept in mind, however, that the numerical tests presented here
mainly serve to confirm the validity of the proposed equations. They do not
prove that the method can be freely used in calculations of the physical
properties of molecules. It is known that even constraints upon bond lengths
can, in certain cases, cause significant perturbations \cite{Watanabe:95},
and that fixing bond angles may significantly affect the conformational
statistics via the so-called metric tensor
\cite{Fixman:74,Fixman:78,vanGunsteren:80}. Effects of each type of
constraint upon the measured properties should, therefore, be thoroughly
studied. These reservations, however, are not that important for many
applications in structural refinement of biopolymers \cite{Rice:94}, in
conformational searches and in simulations of protein folding.

\acknowledgments

This study was initially motivated and partially based upon unpublished
results obtained in collaboration with V. E. Dorofeyev \cite{Dor-phd:}.
I wish to thank Dr. R. Lavery for useful discussions, without his support
this work could have not be finished. I also would like to thank Dr. G.
Rodriguez of the Jet Propulsion Laboratory for making the set of his reprints
available for me in Russia.

\appendix\section*{Implicit Leapfrog Integrator}

To simplify notation let us consider the case of a single variable. Let
$\theta$ and $p$ denote the generalized coordinate and the corresponding
conjugate momentum. Eq. (19) may be rewritten as
\begin{equation}
\dot{p}=g+f
\end{equation}
where $g=dU/d\theta$ and $f$ denotes the inertial term. Calculation
of $g$ may be denoted as $\left(\theta\right)\rightarrow g$, calculation
of $f$ as $\left(\theta,\dot{\theta}\right)\rightarrow f$, etc. The on-step
values are denoted as $\theta_n,\theta_{n+1},...$, etc. and the half-step
values as $\theta_{n-\frac 12},\theta_{n+\frac 12},...$, etc. Then the
sequence of steps in the leapfrog integrator is expressed as
\begin{equation}
\left(\theta_n\right)\rightarrow g_n
\end{equation}
\begin{equation}
p_{n+\frac 12}=p_{n-\frac 12}+\left( g_n+f_{n-\frac 12}\right) h
\end{equation}
\begin{equation}
\theta_{n+\frac 12}=\theta_{n-\frac 12}+h\dot{\theta}_{n-\frac 12}
\end{equation}
\begin{equation}
\left(\theta_{n+\frac 12},p_{n+\frac 12}\right)\rightarrow\dot{\theta}
_{n+\frac 12}
\end{equation}
\begin{equation}
\left\lbrace
 \begin{array}{rclr}
 \theta_{n+\frac 12}&=&\theta_{n-\frac 12}+\left(\dot{\theta}
 _{n-\frac 12}+\dot{\theta}_{n+\frac 12}\right)\frac h2&\ \ (a)\\
 \left(\theta_{n+\frac 12},\dot{\theta}_{n+\frac 12}\right)&\rightarrow&
 f_{n+\frac 12}&\ \ (b)\\
 p_{n+\frac 12}&=&p_{n-\frac 12}+g_nh+\left(f_{n-\frac 12}
 +f_{n+\frac 12}\right)\frac h2&\ \ (c)\\
 \left(\theta _{n+\frac 12},p_{n+\frac 12}\right)&\rightarrow&
 \dot{\theta}_{n+\frac 12}&\ \ (d)\\
 \end{array}
\right.
\end{equation}
\begin{equation}
\theta_{n+1}=\theta_n+\dot\theta_{n+\frac 12}h
\end{equation}
where $h$ denotes the step size.

The steps enclosed in the brace are repeated iteratively until convergence
of Eq. (6A). The iterative cycle involves two stages with non-trivial
calculations given by Eqs (7A) and (9A). The half-step coordinates are
corrected in Eq. (6A) therefore the inversion of the mass matrix is
implicitly involved in Eq.(9A), which makes the algorithm relatively costly
and leaves some room for possible improvements.

Some estimates of the CPU time made in the course of the numerical tests are
given in Table 1A. The measurements were made on a Silicon Graphics Crimson
R4000-50 work station. DY was complied by a standard C-compiler with the
lowest level of optimization. In all calculations a relative tolerance of 10$
^{-6}$ was used as the criterion of convergence of Eq.(6A). The cost of the
matrix inversion calculations scales roughly linearly with the number of
variables therefore it is much less expensive in the standard geometry
calculations. The average number of iterations depends upon the molecular
model and also upon the step size. With the optimal time steps it was
minimal in the standard geometry simulations (3.9) and maximal in
simulations with fixed bond lengths to hydrogens (9.5). Thus the standard
geometry approximation is the cheapest both because of the reduced number of
variables and faster convergence. Evaluation of forces takes similar time in
AMBER and in DY.

\begin{figure}
\caption{Illustration of the definition of external and internal coordinates
in the molecule: $\alpha$, $\beta$, $\gamma$, $\delta$ and $\varepsilon$
denote atom numbers; $\omega$, $\varphi$ and $l$ are internal coordinates
used to position atom $\delta$; $xyz$ and $x^{\prime}y^{\prime}z^{\prime}$
denote the global and the local coordinate frames, respectively. Detailed
explanations are given in the text.}
\end{figure}

\begin{figure}
\caption{Time step dependencies of the total energy drift and the relative
fluctuation for four models of PGB: (a) and (b) - without constraints, (c)
and (d) - with fixed bond lengths to hydrogen atoms, (e) and (f) - in the
standard geometry approximation, (g) and (h) - in the standard geometry
approximation with inertia of hydrogen-only rigid bodies increased as
described in the text. Thinner line - Cartesian coordinate dynamics by
AMBER. Thicker line - internal coordinate dynamics by DY.}
\end{figure}

\begin{table}
\caption{Timing Comparisons. CPU time in seconds per one step
of dynamics.}
\begin{tabular}{lcccc}&A\tablenote{No constraints. Time step 1 fsec.}
&B\tablenote{Fixed bond lengths to hydrogen atoms. Time step 2 fsec.}
&C\tablenote{All bond lengths fixed. Time step 2 fsec.}
&D\tablenote{Standard geometry. Time step 4 fsec.}\\\hline
AMBER&1.10&1.12&1.20&-\\ DY&1.60&2.33&1.97&1.12
\end{tabular}
\end{table}

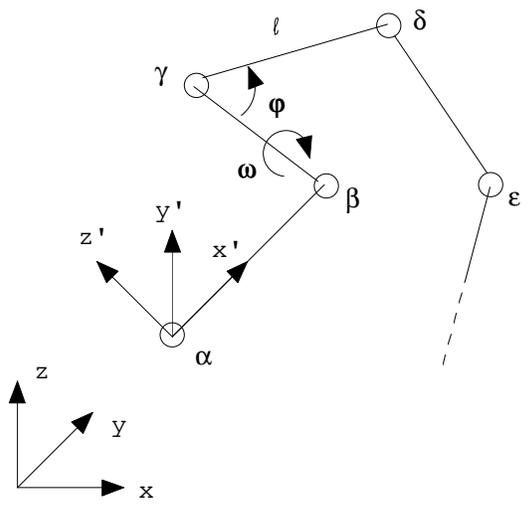
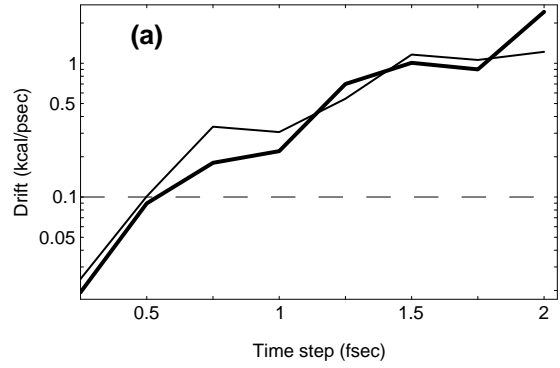
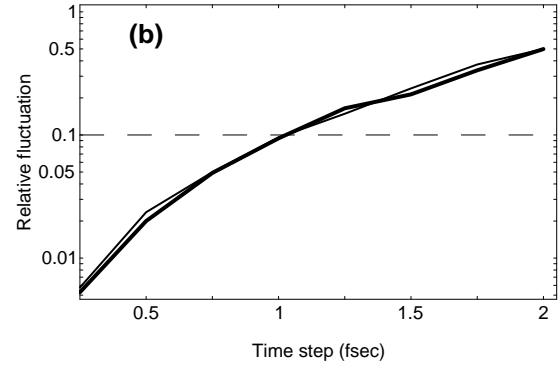
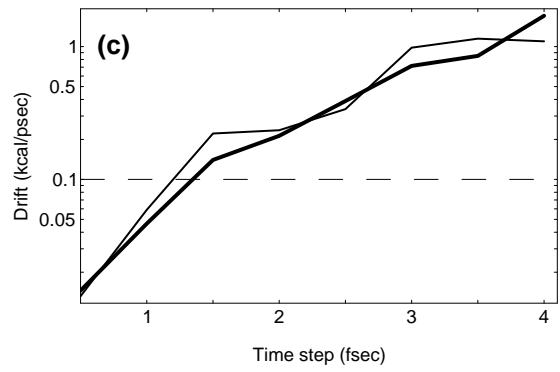
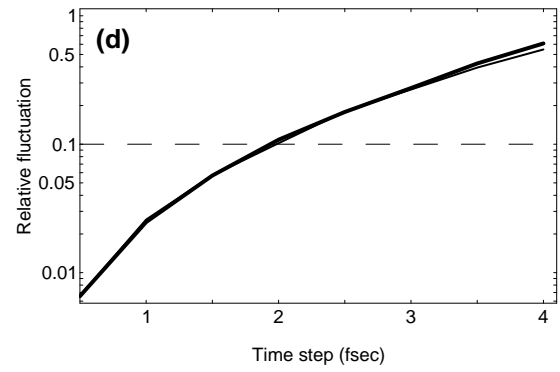
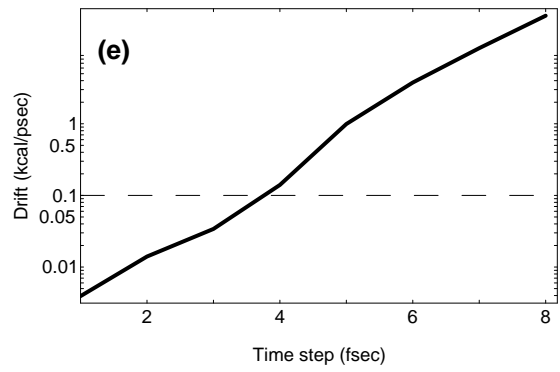
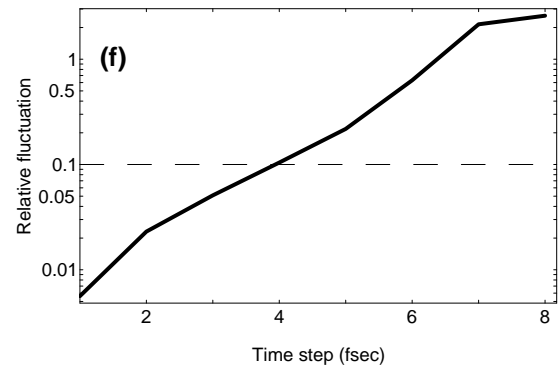
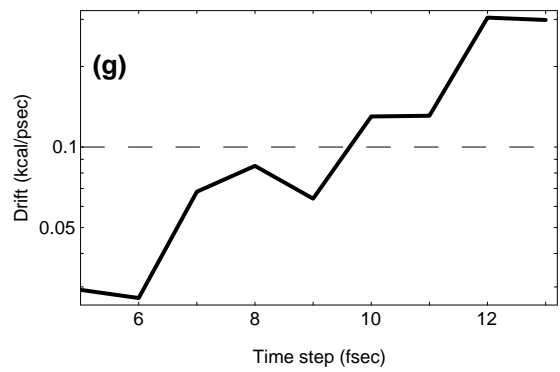
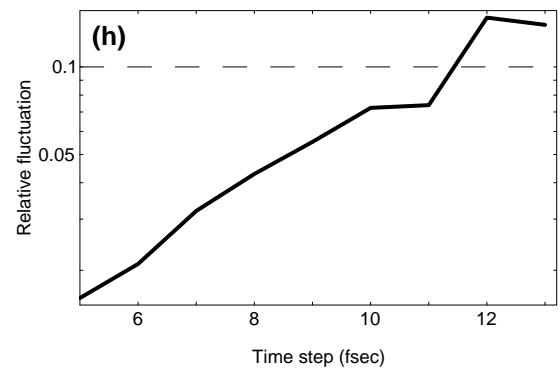
\end{document}